\documentclass[journal]{new-aiaa}
\usepackage[utf8]{inputenc}

\usepackage{graphicx}
\usepackage{amsmath}
\usepackage{setspace}
\usepackage[version=4]{mhchem}
\usepackage{siunitx}
\usepackage{longtable,tabularx}
\usepackage{subcaption}
\usepackage{bm}
\usepackage{xcolor}
\title{Numerical Bow Shock Instabilities in Inert Polyatomic Gases}

\author{G. S. Sidharth\footnote{Assistant Professor. Corresponding author: sidharth\_gs@hotmail.com}}
\affil{Aerospace Engineering, Iowa State University, Ames, IA 50011}
\author{A. Dwivedi \footnote{Research Associate}}
\affil{Aerospace Engineering and Mechanics, University of Minnesota, Minneapolis, MN 55455}

\begin{document}
\doublespacing

\maketitle

\begin{abstract}
We investigate inviscid numerical instabilities that arise in simulations of axisymmetric flow over a hypersonic sphere in an inert, calorically perfect gas at low specific heat ratio ($\gamma \approx 1.1$--$1.2$).
We show that when the density ratio across the bow shock is high and the computational mesh is relatively coarse, numerically induced traveling-wave instabilities of the carbuncle type can develop in the shock layer near stagnation for inert gases. These instabilities, not previously documented in the literature, are noteworthy because bow shock oscillations are also observed experimentally in polyatomic gases exhibiting post-shock thermochemical relaxation. When such gases are modeled as inert with an effectively low $\gamma$, our results emphasize the need for caution to avoid conflating genuine physical instabilities with numerical artifacts in simulations.
\end{abstract}



\section{Introduction}

When a supersonic or hypersonic flow encounters a blunt body, a curved detached bow shock forms upstream of the body. Under certain conditions, the bow shock and the shock layer behind it can become unstable, exhibiting large-amplitude oscillations and unsteady spatial structure. This phenomenon has attracted sustained attention since the early observations of Baryshnikov et al.~\cite{baryshnikov1979}, who reported chaotic perturbations of the shock front and turbulent flow behind it in ballistic-range experiments with the heavy polyatomic refrigerants \ce{CCl2F2} and \ce{CF4}. These gases were selected deliberately. At the impact velocities used, the post-shock temperatures are sufficient to initiate molecular decomposition, and the resulting endothermic dissociation absorbs a large fraction of the thermal energy, driving the post-shock density far above its frozen-chemistry value.

An important parameter for the instability is the normal-shock density ratio $\rho_2/\rho_1$, which in turn depends on both the Mach number and the effective specific heat ratio $\gamma$ of the gas. At a curved shock, misalignment between pressure and density gradients generates vorticity through the baroclinic mechanism. As shown by Hornung and Lemieux~\cite{hornung2001}, the vorticity imparted to a fluid element is proportional to the shock curvature and the density ratio.
\begin{equation}
  \omega \propto U_\infty \kappa \left(1 - \frac{\rho_1}{\rho_2}\right)^2 \frac{\rho_2}{\rho_1}.
\end{equation}
For large density ratios the vorticity is approximately proportional to $\rho_2/\rho_1$, generating a shear layer in the shock layer that can become Kelvin-Helmholtz unstable. Hornung and Lemieux~\cite{hornung2001} identified two nonlinear amplification mechanisms. First, vortical structures in the shear layer move supersonically relative to the gas beneath them, generating oblique shocks that reflect from the body surface and reinforce the structures. Second, once sufficiently amplified, the structures corrugate the bow shock itself, producing triple points that shed secondary shear layers. Their experiments, in which spherical projectiles were fired at high velocity through propane and \ce{CO2} using a light gas gun at Caltech, confirmed a critical density ratio of approximately 14. It must be noted that, at the velocities used, the gases are expected to produce vibrational excitation or dissociation due to large post-shock temperatures.

Baryshnikov and colleagues~\cite{baryshnikov2008,baryshnikov2016,baryshnikov2024} approached the problem from a different angle, focusing on the thermochemical origin of the vorticity. Baryshnikov~\cite{baryshnikov2008} analyzed the spectral characteristics of the turbulent flow behind the bow shock in dissociating \ce{CF2Cl2} and found that the energy-containing vortex scale corresponded directly to the dissociation energy of the gas molecules. By relating the entropy gradient produced by the chemical decomposition to the baroclinic vorticity source term, he derived an instability criterion based on the competition between the chemical dissociation energy and the kinetic energy of the flow. Baryshnikov et al.~\cite{baryshnikov2016} showed that the critical Mach number threshold corresponds precisely to the pressure and temperature conditions required for chemical decomposition of the polyatomic gas, and that the instability is found for different body shapes under the same chemical conditions. Most recently, Baryshnikov et al.~\cite{baryshnikov2024} derived that baroclinic instability of the shock layer can arise from the internal physicochemical energy release alone, independently of boundary conditions on the body, and that the predicted instability development time agreed with measurements in \ce{CF2Cl2}.

Ohnishi et al.~\cite{ohnishi2015} performed three-dimensional numerical simulations and experimental demonstrations using the hydrofluorocarbon HFC-134a ($\gamma = 1.12$) at Mach 9.6, a gas that likewise undergoes significant vibrational relaxation under the relevant post-shock conditions. Their parametric study in $\gamma$--$M_\infty$ space estimated a critical density ratio of approximately 10, somewhat below the value of Hornung and Lemieux due to differences in body geometry. They identified a Helmholtz resonator-like feedback mechanism in which a shear layer from the body edge generates acoustic waves that propagate upstream to perturb the bow shock, and their experiments confirmed the shock deformation predicted numerically. More recently, Alvarez and Lozano-Dur{\'a}n~\cite{anton2025} have studied the global stability of the bow shock to explain instabilities observed in their large eddy simulations \cite{anton2025prf} of Mars re-entry simulations. 

A closely related class of unsteady and unstable shock phenomena arises in shock-induced combustion (SIC), where a blunt projectile travels at hypervelocity through a premixed combustible gas mixture. The classic experiments of Lehr~\cite{lehr1972} showed that, depending on the projectile velocity and mixture, either steady or periodically oscillating flow can result. The theoretical model of McVey and Toong~\cite{mcvey1971} and the detailed numerical studies of Matsuo et al.~\cite{matsuo1995} and Wilson and Sussman~\cite{wilson1993}established that the periodic oscillations of the bow shock and reaction front are governed by the chemical induction time. Wave interactions between compression waves, the shock, and the reaction front generate a limit cycle whose frequency is set by the ratio of the induction length to the projectile radius. Critically, grid-convergence studies in these reactive simulations show that the oscillation frequency changes by only a few percent between meshes differing by a factor of four in resolution~\cite{wilson1993,matsuo1995}, establishing that the characteristic scale is anchored by chemistry, not by the grid.

The distinction between physical instability and numerical artifacts is likely to be more subtle than a density-ratio threshold. Shock-capturing methods are susceptible to the carbuncle phenomenon~\cite{quirk1994,robinet2000}, a spurious multi-dimensional instability that plagues normal shocks.
Several studies have analyzed the source of this numerical instability via stability techniques \cite{dumbser2004,ren2020} and several studies have proposed simple and effective numerical treatments to suppress it \cite{liou2000,maccormack2013,simon2019,rodionov2017,chen2020}.
This is further complicated on stretched meshes, where the truncation error varies spatially across the standoff region and can drive wave-like perturbations. 
Whether such perturbations represent genuine flow physics or grid-induced noise is difficult to determine. There are papers that argue that carbuncle instabilities are physically valid solutions \cite{elling2009,moschetta2001}.
Nevertheless, the existence of a grid-independent characteristic length scale and frequency is a generally necessary for physical instability. 

The present study investigates the nature of numerical instabilities in inviscid simulations of hypersonic flow over a sphere at low $\gamma$ conditions, demonstrating that large density ratios alone can produce grid-driven instabilities that vanish upon mesh refinement or numerical treatment, and that the presence of rate-processes are likely essential for producing the grid-independent scales observed in experiments.
While steady carbuncle instabilities have been observed in prior simulations of hypersonic flow over blunt bodies, the present work is the first to demonstrate well-organized travelling wave patterns that arise purely from numerical artifacts in the absence of chemistry or real-gas effects, and to characterize their dependence on grid resolution.
The results establish a baseline for numerical contributions to shock layer unsteadiness that must be cautiously accounted for in 
this class of flows. 

This article is organized as follows. Section~2 describes the computational setup, including flow conditions, grid design, and the numerical method. Section~3 presents the results: the effect of the eigenvalue limiter on instability character, the grid dependence of instability amplitude, and the distinction between uniform and stretched grids. Section~4 summarizes the conclusions.

\section{Computational Setup}

\subsection{Flow Conditions and Geometry}

\begin{figure}[hbt!]
\centering
\includegraphics[width=0.95\textwidth]{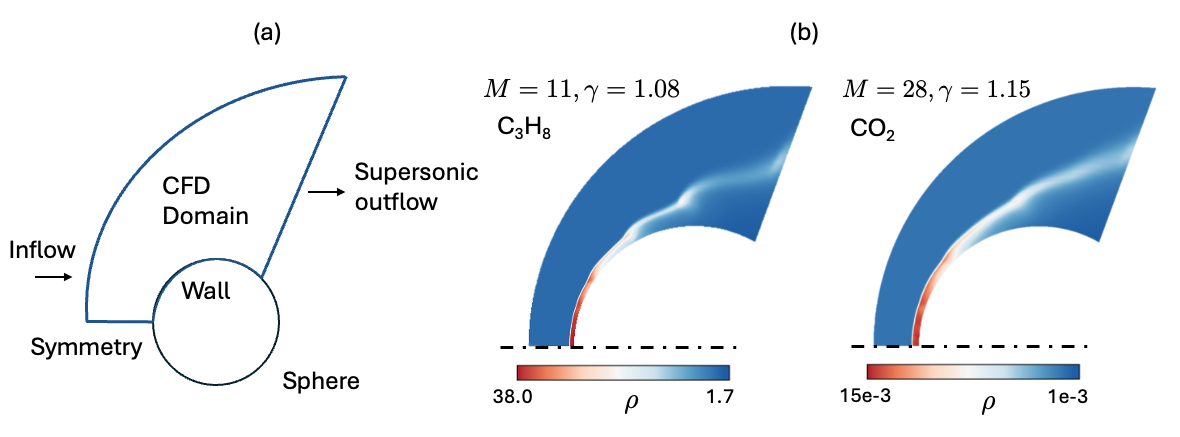}
\caption{(a) Axisymmetric computational domain. (b) Instantaneous density fields for two different examples that exhibit shock-layer corrugations.}
\label{fig:setup}
\end{figure}

We consider two flow configurations, both involving hypersonic flow over a sphere of diameter $D$. The first is representative of the Mars Science Laboratory (MSL) entry at $M_\infty = 28$ in \ce{CO2} ($\gamma \approx 1.15$), which yields a normal-shock density ratio of approximately 14. The second corresponds to conditions similar to those of Hornung and Lemieux~\cite{hornung2001}, with propane (\ce{C3H8}) at $M_\infty = 11$ and $\gamma \approx 1.08$, producing a density ratio exceeding 20. These two cases, shown in Figure~\ref{fig:setup} at a grid spacing $\Delta_s \approx 0.1$ (where $\Delta_s$ is the cell size at the shock normalized by the shock standoff distance, which is approximately $0.018D$), demonstrate that the numerical instabilities are not specific to a single flow condition. They arise generically whenever the density ratio across the bow shock is large, as occurs for polyatomic gases at high Mach numbers. 
All simulations solve the axisymmetric Euler equations unless otherwise noted; selected cases include viscous (Navier-Stokes) computations to assess the role of viscosity.

\subsection{Grid details}

All grids are structured, body-fitted, and stretched in the wall-normal direction. The domain extends to 0.68D at the stagnation line. The circumferential spacing is fixed at $\Delta_\theta = 0.6^\circ$ and the wall-normal spacing at the sphere surface is 1~$\mu$m. We use 60, 100, and 200 points in the wall-normal direction, giving normalized shock-cell spacings of $\Delta_s \approx 0.1$, $0.05$, and $0.025$, respectively; the 60- and 100-point grids are sufficient to demonstrate the instability. No additional clustering is applied near the bow shock, so $\Delta_s$ remains the primary grid-resolution parameter.
While uniform-mesh cases are also computed on grids of comparable nominal resolution in the standoff region, the results of only the stretched grids are presented, because they are relevant for aeroheating simulations. 

\subsection{Numerical Method}

The compressible Euler and Navier-Stokes equations are solved using a finite-volume method with a second-order accurate Roe-type Riemann solver and MUSCL reconstruction. Time integration uses an explicit Runge-Kutta scheme. No artificial dissipation or entropy fix beyond the standard Roe scheme is applied. The solver is a standard shock-capturing method that is susceptible to the carbuncle phenomenon, which is intentional. We wish to characterize the instabilities that arise under typical production conditions rather than under specially stabilized methods.

To motivate the eigenvalue limiter, it is useful to recall the structure of the upwind flux. The convective flux Jacobian $\mathbf{A} = \partial \mathbf{F}/\partial \mathbf{U}$ admits the diagonalization $\mathbf{A} = \mathbf{R}^{-1}\boldsymbol{\Lambda}\mathbf{R}$, where $\mathbf{R}$ and $\mathbf{R}^{-1}$ are the right and left eigenvector matrices and $\boldsymbol{\Lambda} = \mathrm{diag}(\lambda_1,\ldots)$ is the diagonal matrix of eigenvalues \cite{candler2015}. For the three-dimensional Euler equations, the eigenvalues are
\begin{equation}
  \lambda = u', \quad \lambda^{\pm} = u' \pm a,
  \label{eq:eigs}
\end{equation}
where $u' = \mathbf{u}\cdot\hat{n}$ is the face-normal velocity component and $a$ is the speed of sound; $\lambda$ appears with multiplicity three (corresponding to the entropy and transverse-momentum waves) while $\lambda^{\pm}$ are the acoustic waves. Using the splitting $\boldsymbol{\Lambda}^{\pm} = \tfrac{1}{2}(\boldsymbol{\Lambda} \pm |\boldsymbol{\Lambda}|)$, the modified Steger--Warming and Roe fluxes can both be written in the form \cite{candler2015}
\begin{equation}
  \mathbf{F}_{i+1/2} = \tfrac{1}{2}\bigl(\mathbf{F}_{i+1}+\mathbf{F}_i\bigr)
   - \tfrac{1}{2}\bigl(\mathbf{R}^{-1}|\boldsymbol{\Lambda}|\mathbf{R}\bigr)_{i+1/2}\bigl(\mathbf{U}_{i+1}-\mathbf{U}_i\bigr)
  \label{eq:flux}
\end{equation}
making explicit that the amount of upwind dissipation applied to each characteristic wave is controlled by the magnitude of the corresponding eigenvalue $|\lambda|$.
A well-known stabilization technique in high-speed CFD is eigenvalue limiting, in which the eigenvalues in the dissipative term of \eqref{eq:flux} are prevented from approaching zero. Following \cite{candler2015}, this is accomplished by replacing each eigenvalue $\lambda$ with
\begin{equation}
  \lambda' = \frac{1}{2}\!\left(\lambda \pm \sqrt{\lambda^2 + \varepsilon^2}\right),
  \label{eq:eiglim}
\end{equation}
where the sign is chosen according to whether $\lambda$ contributes to $\boldsymbol{\Lambda}^+$ or $\boldsymbol{\Lambda}^-$, and the limiting scale is $\varepsilon = \varepsilon_0 (|u'| + a)$, with $\varepsilon_0$ typically in the range $0.1 \leq \varepsilon_0 \leq 0.3$. As $u' \to 0$ near a stagnation point, the entropy-wave eigenvalue $\lambda = u'$ in \eqref{eq:eigs} vanishes, so $|\lambda| \to 0$ in the dissipative term of \eqref{eq:flux}, and upwind dissipation is extinguished precisely where the normal shock is located, triggering carbuncle-type error accumulation. Equation~\eqref{eq:eiglim} enforces a floor on $|\lambda|$ of order $\varepsilon_0 a$, restoring numerical stability in the stagnation region. This limiter is closely related to the entropy fix and sonic-glitch corrections used at sonic points in upwind schemes. Importantly, it is not applied inside boundary layers, where the same artificial inflation of $|\lambda|$ can introduce a spurious cross-layer flux \cite{candler2015}.


\section{Results}

\subsection{Flow conditions in a low Specific Heat Ratio gas}

To orient the reader, we first discuss the qualitative change in flow behavior as $\gamma$ decreases. Let us consider the propane case, in which $\gamma = 1.08$ at $M_\infty = 11$. Comparing this to the bow shock at $\gamma = 1.4$, the density ratio across the normal shock increases from 5.76 to 21.5 when $\gamma$ is reduced to 1.08. 
The shock standoff distance, being inversely proportional to the density jump, decreases from approximately $0.02D$ to $0.015D$, and the shock curvature at the stagnation point increases. The combination of higher density ratio and stronger curvature leads to significantly larger vorticity in the shock layer, which is susceptible to the usual shear-layer instabilities that corrugate the bow shock and produce triple points. 
We now discuss the solution with an eigenvalue limiter of value $\varepsilon_0 = 0.3$ and a grid with $N_y = 60$ points in the wall-normal direction.  
For $\gamma = 1.4$, the flow field is steady. The bow shock remains axisymmetric and the shock layer exhibits no unsteadiness, despite low resolution at the shock. 
As $\gamma$ is reduced to the value of $\gamma \approx 1.08$, the bow shock standoff distance decreases substantially and the shock is better resolved on the same grid, with $\Delta_s \approx 0.1$. Under these conditions, numerical instabilities emerge such that the bow shock develops visible corrugations and the shock layer exhibits unsteady traveling wave patterns that propagate along the bow shock in the circumferential direction.

\subsection{Effect of the eigenvalue limiter}

\begin{figure}[hbt!]
\centering
\includegraphics[width=0.80\textwidth]{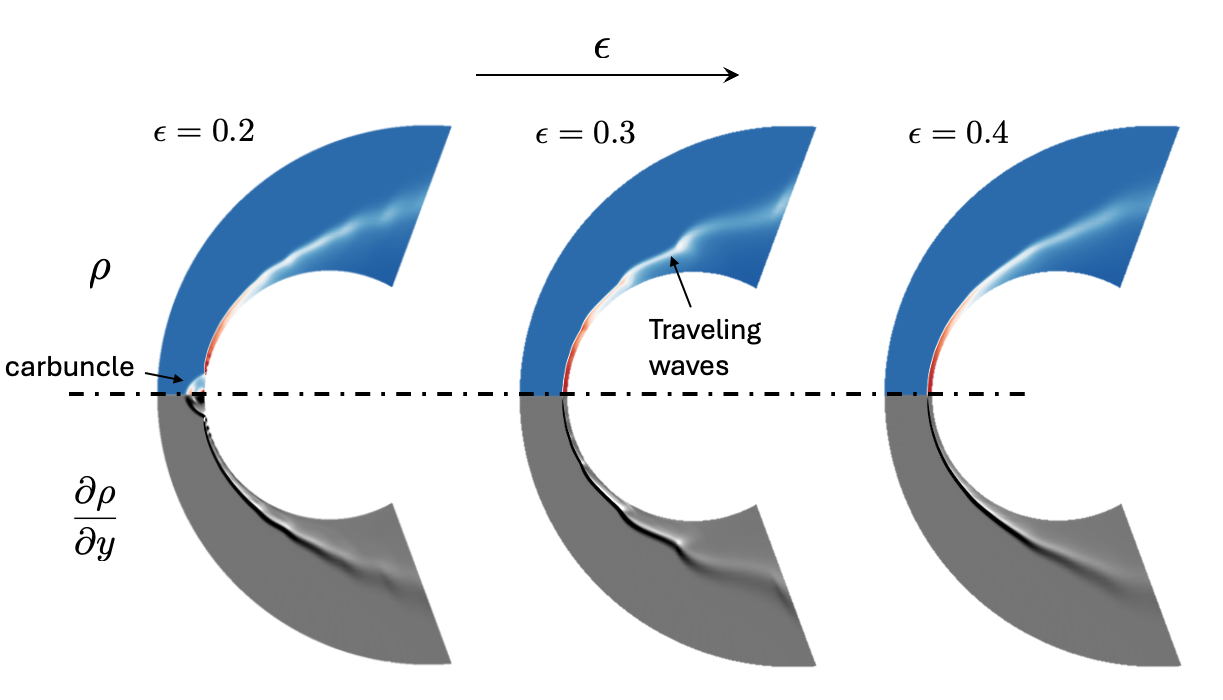}
\caption{Effect of the eigenvalue limiter parameter $\varepsilon$ on the shock-layer instability. Top row shows instantaneous density $\rho$; bottom row shows wall-normal density gradient $\partial\rho/\partial y$.}
\label{fig:eps}
\end{figure}

We first examine the effect of the eigenvalue limiter on the shock-layer instability. The limiter is applied to the entropy-wave eigenvalue, which vanishes at the stagnation point and is responsible for the carbuncle instability. By increasing $\varepsilon_0$, we can suppress the carbuncle and observe how the instability evolves. 
Figure~\ref{fig:eps} shows the instantaneous density and wall-normal density gradient for three values of $\varepsilon_0$ (0.2, 0.3, and 0.4). At $\varepsilon = 0.2$, a steady carbuncle deformation is present at the stagnation point, characterized by a localized kink in the bow shock and a corresponding density perturbation in the shock layer. As $\varepsilon$ is increased to $0.3$, the instability transitions to a travelling-wave regime, with wave-like perturbations propagating circumferentially along the shock layer. At $\varepsilon = 0.4$, the shock layer is largely stabilized, with only small residual perturbations remaining. This behavior confirms that the bow shock instabilities can be controlled through numerical treatment generally used to treat carbuncles.

While the mechanisms of amplification may be physical because of the shear layer as has been pointed out by Hornung and Lemieux, this seems to be an intermediate stage between a steady carbuncle and a fully stable bow shock. 

\subsection{Effect of Grid Resolution}

The instability is strongly dependent on grid resolution. On the coarsest grid with $\Delta_s \approx 0.1$, the instability is most pronounced at low values of the eigenvalue limiter $\varepsilon$. 
When the grid resolution is increased to $\Delta_s \approx 0.05$ with $N_y=100$, the instability amplitude decreases significantly, and the shock layer is more stable even at lower $\varepsilon$ values. This suggests that the perturbations that destabilize to produce shock corrugations are numerical artifacts and may be mitigated through grid refinement, provided the carbuncle treatment exists in the numerical method.

For the finer grid the instability amplitude drops by several orders of magnitude as $\varepsilon$ increases, confirming suppression by the eigenvalue limiter. The coarse grid remains unstable across all $\varepsilon$ values tested.

\begin{figure}[hbt!]
\centering
\includegraphics[width=0.65\textwidth]{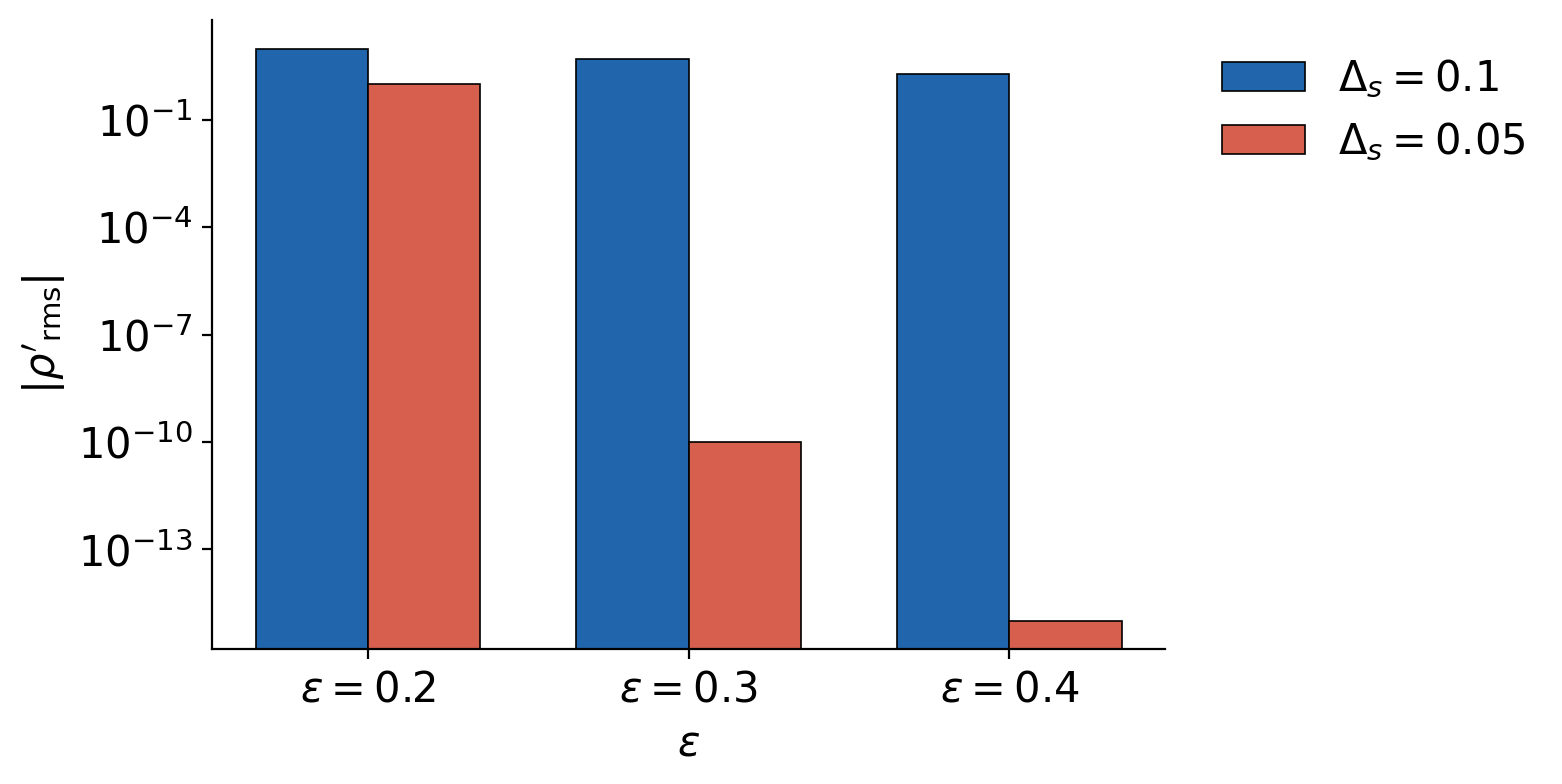}
\caption{RMS density perturbation amplitude $|\rho'_\mathrm{rms}|$ as a function of eigenvalue limiter parameter $\varepsilon$ for two grid spacings ($\Delta_s = 0.1$ and $\Delta_s = 0.05$).}
\label{fig:ampl}
\end{figure}

Figure \ref{fig:ampl} quantifies the root-mean-square (RMS) amplitude of the density perturbations in the shock layer as a function of $\varepsilon$ for both grid resolutions. For the finer grid, the instability amplitude drops by several orders of magnitude as $\varepsilon$ increases, confirming that the eigenvalue limiter effectively suppresses the numerical instability. As seen in the previous figure, the coarse grid, while exhibiting improvement in the bow shock corrugations, remains unstable across all $\varepsilon$ values tested, indicating that grid resolution is an important factor in inciting/controlling the instability.

\begin{figure}[hbt!]
\centering
\includegraphics[width=0.95\textwidth]{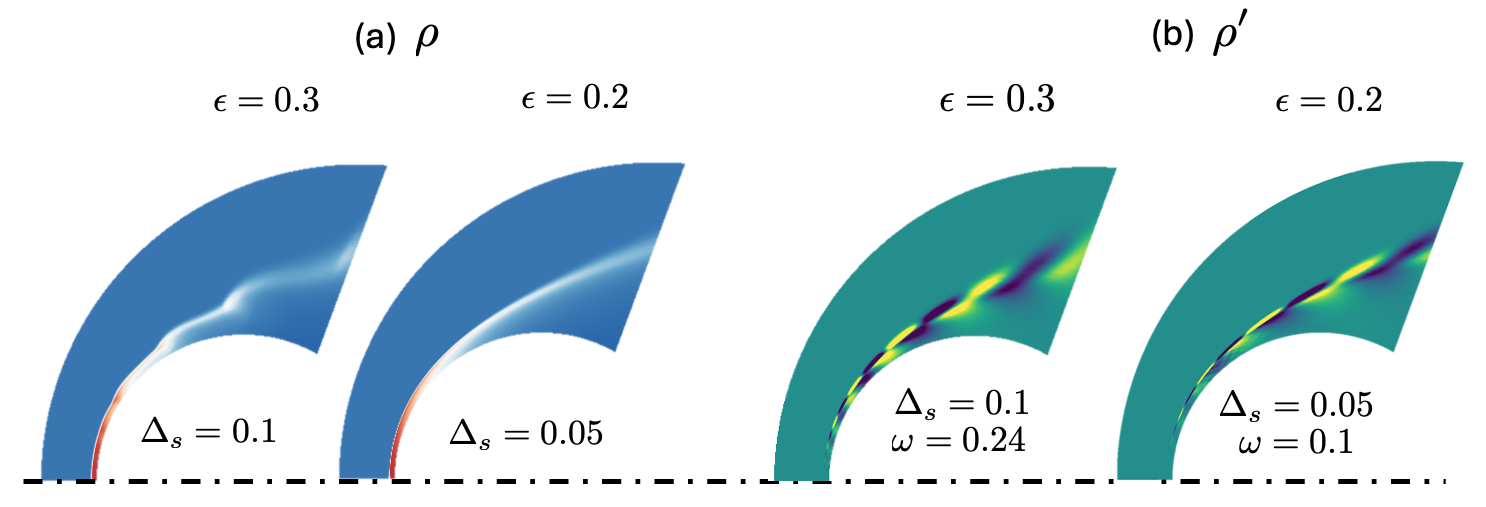}
\caption{Spatial structure of the instability for two cases ($\Delta_s=0.1$, $\varepsilon=0.3$) and ($\Delta_s=0.05$, $\varepsilon=0.2$): (a) instantaneous density $\rho$ and (b) density perturbation $\rho'$ of the dominant SPOD mode.}
\label{fig:modes}
\end{figure}

In these particular examples, the instability manifests as a travelling wave pattern that propagates circumferentially along the shock layer. The wave amplitude and wavelength are both influenced by the numerical resolution, with finer grids producing smaller-amplitude, shorter-wavelength waves, and coarser grids producing larger-amplitude, longer-wavelength waves.
The instability exhibits a characteristic wavelength $\lambda$ that scales with the sphere radius. Figure~\ref{fig:modes} shows the extracted instability spectral proper orthogonal modes \cite{he2021,towne2018} at two grid resolutions and $\varepsilon$ values for the propane case. We show the comparison with a lower value of $\varepsilon$ in the finer grid because at the same value of $\varepsilon$, the signature of the instability vanishes and only shock-grid misalignment errors persist. At these two values however, the spatial structure of the traveling wave instability is similar but the frequency reduces by approximately a factor of 2 in the finer grid, with a clear indication of numerics-seeded unstable waves.



\subsection{Inviscid versus Viscous Calculations}

To assess whether boundary layers or other viscous effects can stabilize or suppress the numerical instability, we perform viscous (Navier-Stokes) simulations on the same stretched grids at Reynolds numbers relevant to the ballistic-range conditions studied by Hornung and Lemieux~\cite{hornung2001}. The boundary layer is resolved on the body surface, but the standoff region retains the same coarse stretched grid as in the inviscid cases.

The viscous calculations show no significant change in the instability behavior: the shock layer oscillations persist with essentially the same wavelength and amplitude as in the inviscid calculations. This result confirms that the instability is fundamentally inviscid in character, arising from the interaction between large density gradients, shock curvature and the coarse mesh in the standoff region, rather than from any near-wall boundary-layer or viscous instability mechanisms.

\subsection{Uniform versus Stretched Grids}

%
Numerical studies of the carbuncle phenomenon typically employ uniform grids, since the instability is inviscid in nature and uniform grids simplify the analysis. In practice, however, aerothermodynamic simulations require stretched grids to resolve the stagnation-point boundary layer, and we find that grid stretching qualitatively changes the character of the instability. On a uniform grid of comparable nominal resolution, the flow reaches a steady state with a localized kink in the bow shock at the stagnation point but without any propagating disturbances. On a stretched grid, the spatial variation in cell size across the standoff region introduces an inhomogeneity that supports the propagation of grid-induced perturbations. The varying local truncation error acts as a spatially varying forcing that drives the travelling-wave patterns seen in Figure~\ref{fig:modes}.

It is expected that local clustering near the bow shock will introduce additional control variables that affect the numerical instabilities; however, we do not apply any such clustering in the present study. It is worth noting that the presence of clustering near the shock can suppress the travelling wave patterns by reducing the local truncation error and thus the forcing that drives them. However, the coarsening away from the shock may also amplify the instability and is therefore the subject of a future study. 




\section{Conclusion}

This short paper brings to notice that axisymmetric bow shock instabilities can arise from grid and numerical artifacts. While this is not generally a new conclusion, the appearance of numerically generated traveling waves in the absence of chemistry or real-gas effects has not been previously documented, and the grid-scaling behavior of these instabilities has not been characterized.

More specifically, we point out that standard shock-capturing methods on stretched grids can produce grid-artifact instabilities in flows with large density ratios, in the absence of reactive chemistry or real-gas effects. The key requirements are a large normal-shock density ratio (equivalently, a low effective $\gamma$) and coarse grid spacing at the shock.
A caveat is that such inert gas flows may not be physically realizable, given the active molecular rate processes that occur under the relevant post-shock conditions. 
It is important to clarify that the present results do not imply that perturbation amplification mechanisms in curved shocks are unphysical; rather, they highlight numerical seeds or contributions to shock layer unsteadiness that must be carefully accounted for in simulations.
In summary,

\begin{enumerate}
  \item On stretched grids, low $\gamma$ can support travelling wave instabilities whose amplitude increases with the local grid spacing in the standoff region.
  \item The carbuncle treatment of the numerical method at the shock suppresses the instability.
  \item Viscous effects or the presence of a boundary layer does not seem to affect the instability; it is fundamentally an inviscid, curved-shock, grid-driven phenomenon.
\end{enumerate}

An important implication of this work is that the presence of curved shock-layer corrugations and oscillations in a simulation does not by itself confirm that the instability is physical. The key diagnostic is perhaps the grid convergence of the oscillation frequency. In shock-induced combustion simulations~\cite{matsuo1995,wilson1993,sussman1995}, the frequency changes by only a few percent between meshes differing by a factor of four in resolution, anchored by the chemical induction length. In the present calorically perfect gas simulations there is no such anchor, and the instability wavelength tracks the grid and the oscillations vanish upon refinement. 
 Grid convergence studies and, where possible, shock-fitting methods~\cite{anton2025,robinet2000} are necessary to isolate the physical contribution. Notably, in each of the physically documented cases~\cite{lehr1972,hornung2001,baryshnikov2008}, the shock layer is simultaneously a site of active molecular decomposition or combustion, suggesting that such rate processes may play a role beyond merely raising the density ratio via a surrogate $\gamma$. Future work will pursue adjoint-based sensitivity analysis to identify the wavemaker of these instabilities and to develop grid-adaptive strategies that suppress the numerical component without damping physical phenomena.

\section*{Acknowledgments}

The authors acknowledge communication with Prof. H. Hornung for insight into curved shock instabilities and whose experimental data inspired the present investigations. 

\bibliography{refs}

\end{document}